\documentclass[conference]{IEEEtran}
\usepackage{cite}
\usepackage{amsmath,amssymb,amsfonts}
\usepackage{algorithmic}
\usepackage{graphicx}
\usepackage{textcomp}
\usepackage{xcolor}
\def\BibTeX{{\rm B\kern-.05em{\sc i\kern-.025em b}\kern-.08em
    T\kern-.1667em\lower.7ex\hbox{E}\kern-.125emX}}

\usepackage{tikz}
\usepackage{textcomp}
\usepackage{hyperref}
\usepackage{lipsum}
\newcommand\copyrighttext{
  \footnotesize \textcopyright 2019 IEEE.  Personal use of this material is permitted.  Permission from IEEE must be obtained for all other uses, in any current or future media, including reprinting/republishing this material for advertising or promotional purposes, creating new collective works, for resale or redistribution to servers or lists, or reuse of any copyrighted component of this work in other works. 
  
  To be published in the 10th International Conference on Information, Intelligence, Systems and Applications (IISA 2019)
}
\newcommand\copyrightnotice{%
\begin{tikzpicture}[remember picture,overlay]
\node[anchor=south,yshift=10pt] at (current page.south) {\fbox{\parbox{\dimexpr\textwidth-\fboxsep-\fboxrule\relax}{\copyrighttext}}};
\end{tikzpicture}%
}

\begin{document}

\title{A Methodology for Saving Energy in Educational Buildings Using an IoT Infrastructure}

\author{\IEEEauthorblockN{Georgios Mylonas, Dimitrios Amaxilatis, Stelios Tsampas, Lidia Pocero}
\IEEEauthorblockA{
\textit{Computer Technology Institute and Press ``Diophantus''}\\
Patras, 26504, Greece \\
\{mylonasg, amaxilat, tsampas, pocero\}@cti.gr}
\and
\IEEEauthorblockN{Joakim Gunneriusson}
\IEEEauthorblockA{\textit{Staffangymnasiet School}\\
S\"{o}derhamn, Sweden\\
joakim.gunneriusson@hufb.se}
}

\maketitle
\copyrightnotice

\begin{abstract}
A considerable part of recent research in smart cities and IoT has focused on achieving energy savings in buildings and supporting aspects related to sustainability. In this context, the educational community is one of the most important ones to consider, since school buildings constitute a large part of non-residential buildings, while also educating students on sustainability matters is an investment for the future. In this work, we discuss a methodology for achieving energy savings in schools based on the utilization of data produced by an IoT infrastructure installed inside school buildings and related educational scenarios. We present the steps comprising this methodology in detail, along with a set of tangible results achieved within the GAIA project. We also showcase how an IoT infrastructure can support activities in an educational setting and produce concrete outcomes, with typical levels of 20\% energy savings. 
\end{abstract}


\section{Introduction}

The Internet of Things (IoT) and smart cities are two of the most popular directions the research community is currently very actively pursuing. During recent years, a considerable amount of resources has been invested into building related infrastructures, leading to the creation and availability of large-scale smart city and IoT installations around the world. However, there is still an important discussion in progress: how can we utilize such smart city and IoT developments, in order to produce reliable, economically sustainable and socially fair solutions that create public value. 

One path towards these goals is to apply such tools in an educational context and focus on sustainability, while achieving energy savings. In general, the educational community is very important both in terms of size and of significance with respect to our future. Today's students are the citizens of tomorrow, and they should possess the scientific and technological skills to respond to future challenges, such as climate change. Moreover, it is now widely accepted that environmental education is closely linked to citizenship education, a key aspect of the learning process that focuses on transforming students into well-informed, considerate and responsible citizens, helpful in the communities they participate in, be it schools, neighborhoods or the wider world. Ideally, this process should link in-school teaching with issues such as environmental protection. It is a means through which Europe can meet its goals, by equipping citizens, enterprise and industry in Europe with the skills and competences  needed to provide sustainable and competitive solutions to the arising challenges~\cite{science-europe}.

Green Awareness In Action -" GAIA~\cite{gaia-site}, a Horizon 2020 EC-funded project, is developing an IoT platform that combines sensing, web-based tools and gamification elements, in order to address the educational community. Its aim is to increase awareness about energy consumption and sustainability, based on real-world sensor data produced by the school buildings where students and teachers live and work, while also lead towards behavior change in terms of energy efficiency. In this context, we believe an approach based on open-source, replicable and widely available technology, providing a ``foundation'' that allows educators to adapt to the needs of each class, opens up many possibilities.

In this paper, we present a methodology used by the schools participating in GAIA as a way to structure their energy-saving interventions in a way that can be easily implemented, verified and reported, using the data made available by the project to these schools. Our focus on educational communities essentially aims to provide better opportunities for teachers to teach such important aspects through hands-on activities. We first present briefly the overall approach of the project and its infrastructure, and then present the steps of the methodology in more detail. A set of results produced by applying the methodology show that is feasible to have substantial energy savings when supplying schools with real-world data to monitor in real-time the effect of applying energy-saving strategies.

Regarding the structure of this paper, we continue with a discussion on previous related work in Section 2, followed by a short general discussion of the GAIA project and its toolset. We present our methodology in Section 4, and we continue with a presentation of some results from energy saving activities conducted in schools. We conclude this paper and briefly outline our future work in Section 6.

\section{Previous Work}

The European Union is placing a strong focus on energy efficiency with initiatives like Build Up~\cite{build-up}, a portal for energy efficiency in buildings. Overall, the percentage of school buildings among non-residential ones in Europe is around 17\%~\cite{eurostat12}. There is a lot of work in the smart cities domain and energy consumption, such as smart grid and energy disaggregation algorithms. Recent works using the most popular type of monitoring for energy disaggregation, non-intrusive load monitoring (NILM), are surveyed in detail in \cite{nilm-review, gluhak-nilm}. \cite{borgstein-energy-performance-review} discusses recent approaches for benchmarking energy in non-residential buildings, while \cite{building-data-genome} presents a collection of performance analysis tools and algorithm benchmarking for non-residential buildings.

Regarding the current state of the art in inclusion of sustainability and other related aspects in the educational domain, there is a lot of activity taking place with respect to inclusion of makerspace elements in school curricula, aided by the availability of IoT hardware as well. \cite{PAPAVLASOPOULOU201757} summarizes recent activity within the Maker Movement approach, presenting relevant recent findings and open issues in related research. \cite{ERIKSSON20189} discusses a study stemming from a large-scale national testbed in Sweden in schools related to the maker movement, along with the inclusion of maker elements into the school curriculum of Sweden. 

Furthermore, there is a large number of research projects and activities that focus specifically on the energy efficiency domain. In addition, projects like ZEMedS~\cite{zemeds} and School of the Future~\cite{school-future} involved the educational community and energy efficiency. There is also a number of recent projects, like SEACS~\cite{seacs}, that produced material related to sustainability and STEM, or current ones like UMI-Sci-Ed~\cite{umi-sci-ed}, that tie educational content to the Internet of Things. Other recent projects like Charged~\cite{charged-sensors} and Entropy~\cite{entropy} target diverse end-user communities and do not focus on the educational community. 

Regarding our own previous work, aspects of the educational activities and their implementation and integration in the schools participating in the GAIA project are presented in \cite{gaia-ieee-pervasive}, while hands-on lab activities are discussed in \cite{ijcci}. The design and implementation of the cloud-based aspects of the project are discussed in \cite{gaia-sensors}, while the open source hardware aspects of the project are discussed in detail in \cite{hardwareX}.

\section{GAIA and Saving Energy in the Educational Domain}

In this section, we provide a brief overview of the infrastructure of the GAIA project, that is used to verify the effectiveness of the proposed methodology, along with the toolset used in the project as an end-user interface to the measurements produced by this infrastructure. As mentioned in the introduction, the type of infrastructure that can be used to provide data for energy-saving activities is quite generic and the results described here are not tied specifically to the infrastructure utilized in our work.

\subsection{The GAIA Infrastructure}

The real-world IoT deployment developed through the GAIA project provides real-time monitoring of 23 school buildings spread in 3 countries (Greece, Italy and Sweden). Of these buildings, the ones in Greece (19 in total) are situated in different local climatic conditions (suburban and rural areas, small islands, city centers). The year of construction of these buildings ranges from 1950 to 2000. Almost all educational levels (primary-high school) and different profiles (curricula, organization, building characteristics, regulations) are covered. The vast majority of the rooms monitored are used for teaching. 

\begin{figure*}[tb]
\centering
\includegraphics[width=0.95\textwidth]{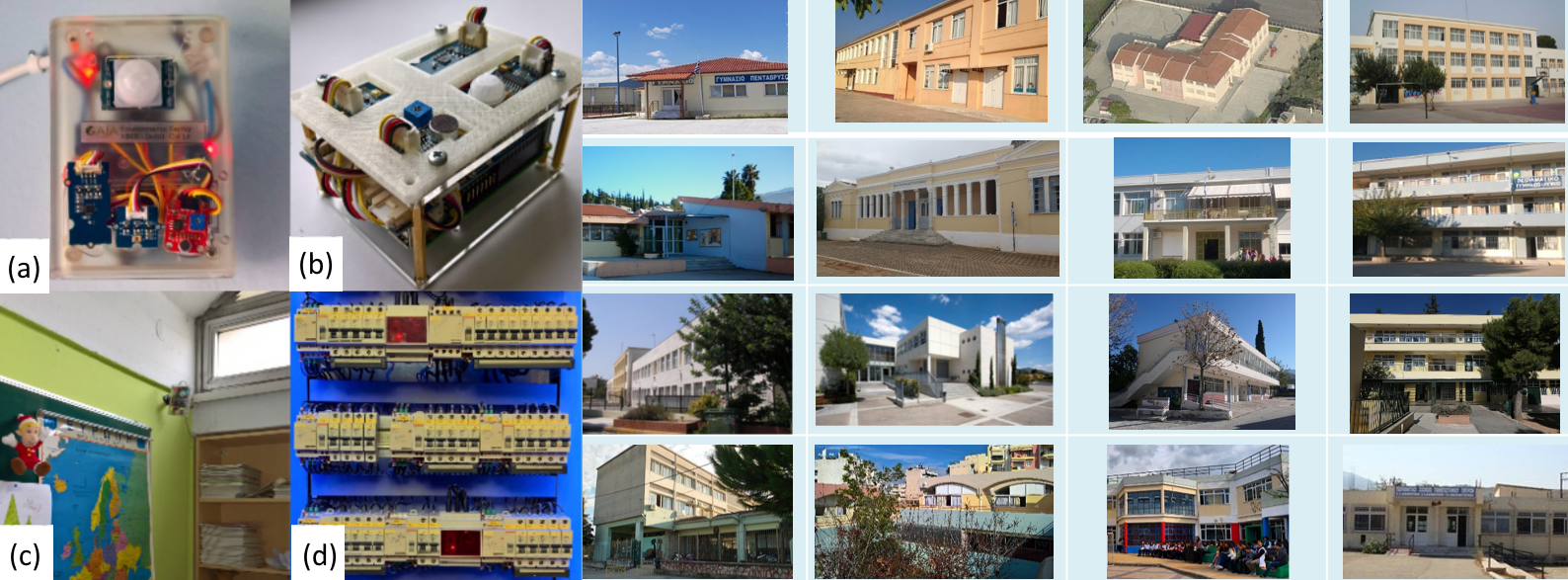}
\caption{Examples of the IoT infrastructure located inside school buildings in Greece (a-b) IoT nodes based on Arduino and Raspberry Pi, c) actual node inside a classroom, d) a power meter installed inside a distribution board at a Greek school, along with photos from the exterior of some of the school buildings.}
\label{fig:gaia-schools}
\end{figure*}

Overall, the deployed devices provide 1250 sensing points organized in four categories: (1) classroom environmental sensors; (2) atmospheric sensors (outdoors); (3) weather stations (on rooftops); and (4) power consumption meters (attached to main building electricity panels). Given the diverse building characteristics and usage requirements, deployments vary between schools (e.g., number of sensors, manufacturer, networking, etc.). The IoT devices (Fig.~\ref{fig:gaia-schools}) used are either open-design IoT nodes~\cite{hardwareX}, or off-the-shelf products from IoT device manufacturers. Indoor devices use IEEE 802.15.4 or LoRa wireless networks. These devices are connected to cloud services via IoT gateway devices, which coordinate communication with the rest of the platform, while outdoors nodes use wired networking or WiFi.

\begin{figure}[htbp]
\centerline{\includegraphics[width=0.85\columnwidth]{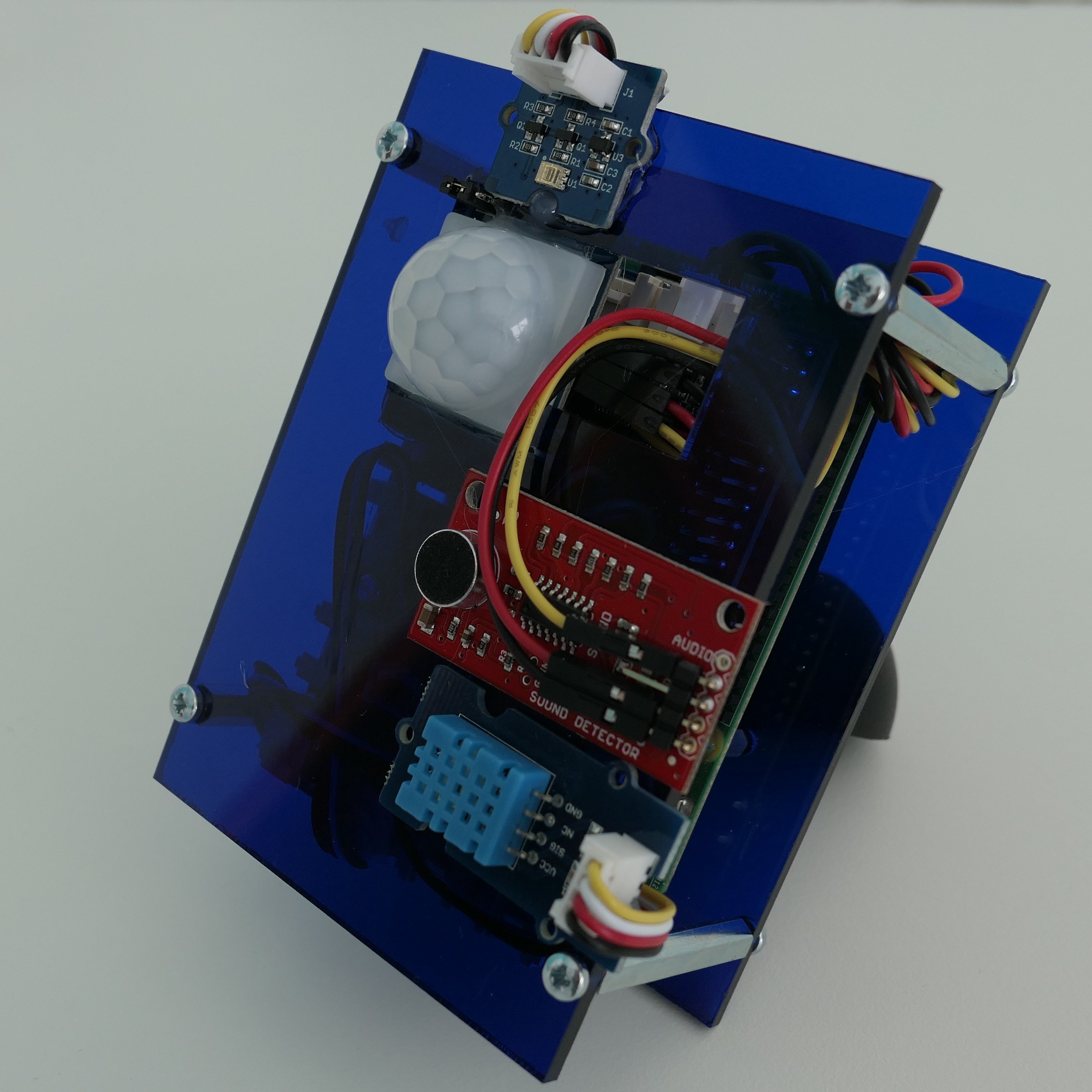}}
\caption{The latest revision of the IoT node used inside classrooms.}
\label{fig:bms}
\end{figure}

\subsection{Translating Data into Action}

This infrastructure would not be particularly useful without having a set of tools to ease the access to the data produced and provide functionality to support educational activities. 

The GAIA Challenge is an online serious game for students, designed to raise energy awareness within their own facility. In addition, real-time data from sensors in the buildings and participatory sensing are a part of the challenge; the aim is to visualize the real-life impact of the participants' behavior and build collaborative (within a facility) and competitive (between facilities) gamification elements upon the real-life impact. The game challenge utilizes gamification mechanics to a) motivate participants to engage in energy saving topics, b) work on online ``quests'', and c) compete and compare against other classes and schools in other countries. It is also localized in a number of languages, and includes diverse activities completed either on individual basis, or a class/group basis. Teachers are involved in some class activities, working together with their students on hands-on observation tasks in classrooms.

The GAIA building manager application is essentially a responsive web-application offering direct visualization of energy consumption and environmental sensing data, while also utilizing participatory sensing in certain scenarios. End-users use it as a means to monitor their school's building status and monitor building performance, i.e., it offers certain building analytics. With respect to building inspection and monitoring, the end-users are able to inspect real-time energy usage where respective meters are available in various timescales (from several minutes to yearly), as well as make comparisons with similar buildings or with the same building in other time spans (e.g., previous years). 

\begin{figure}[htbp]
\centerline{\includegraphics[width=0.98\columnwidth]{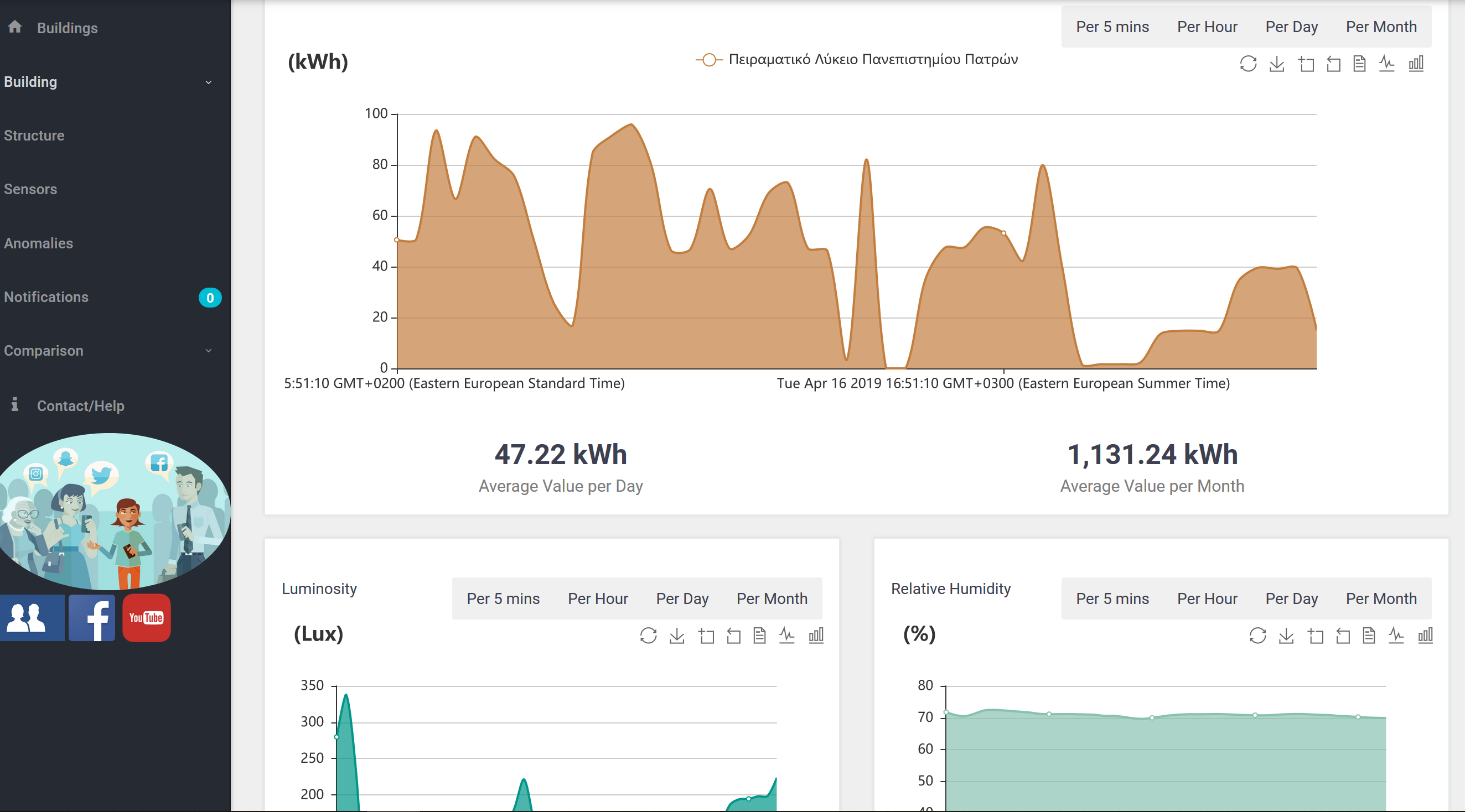}}
\caption{Screenshot from the GAIA BMS Web application.}
\label{fig:bms}
\end{figure}

The Android GAIA Companion app allows end-users with an Android smartphone to access school building data from the GAIA infrastructure in a more immediate manner. Although it does not have the range of visualization options offered by the online building manager application, it provides a faster route to such data, simplifying the implementation of energy saving actions requiring immediate feedback, e.g., determining the energy impact of turning on/off the lights of a specific classroom.

\begin{figure}[htbp]
\centerline{\includegraphics[width=0.95\columnwidth]{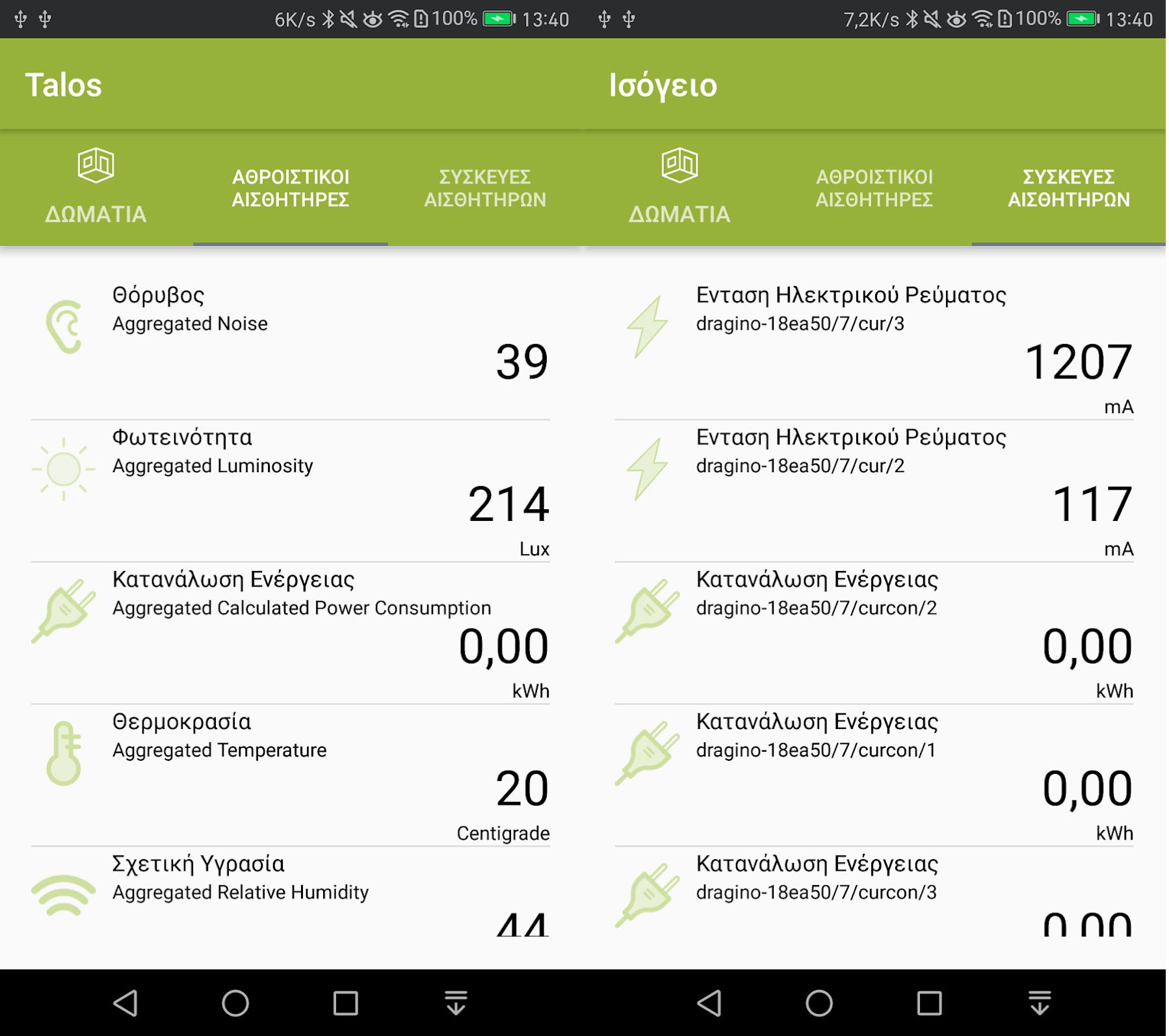}}
\caption{Screenshots from the GAIA Companion Android app.}
\label{fig:gaia-app}
\end{figure}

These software components are used in the context of a set of template activities, that are proposed to the teachers of the schools that participate in the GAIA project. Schools choose the energy-related domain on which they will focus on, and then use the methodology as a way to structure their interventions and be able to monitor in a structured manner. The provided software tools allow for immediate feedback with respect to the effect of the energy-saving strategies they choose to follow and apply inside their schools, which could either belong to the ones proposed by GAIA ,or be something entirely different, e.g., a strategy proposed by the students themselves.

\section{The GAIA Methodology}

In this section, we provide a detailed description of the methodology we propose for integrating energy saving activities into the daily life of a school that has installed an IoT infrastructure in its building to monitor certain parameters, such as its overall power consumption. Its design follows the overall philosophy of the GAIA project, but is not limited to GAIA's implementation or specific hardware/software used in the project. In general, in order to change the behavior of students and teachers in terms of energy consumption and achieve sustainable results, GAIA utilizes a loop-based approach focused around three pillars: raise awareness, support action, and foster engagement. In the context of the proposed methodology, this could be realized by following a series of simple steps, in which students and teachers successively study their environment, monitor the current situation and detect potential issues, devise a strategy to achieve energy savings and act, and then monitor and review the results of their actions. We first include a set of overall guidelines, followed by 2 examples of implementation in schools in Sweden and  Italy. The first one tackled electricity consumption overall in the building, while the second one focused on lights in the school building's corridors.

We now continue with the basic steps in the sample activity implementation methodology: awareness, observation, experimentation and action. A final step suggests staying focused and monitoring progress.

\subsection{Step 1 -" Awareness and Preparatory steps}

This step can be done in parallel with Step 2, or before Step 2. Schools should create a general profile for their building and locate the points where energy is consumed:

\begin{itemize}
\item Lighting inside the building, classrooms and corridors, as well as outside the building. 
\item Heating and air-conditioning.
\item Electrical appliances, e.g., water heating devices, ovens and refrigerators.
\item Equipment used for teaching purposes like PCs, lab equipment, smart boards, 3D printers, etc.
\end{itemize}

The timetable of the school regarding the following aspects should be noted down:

\begin{itemize}
\item Days and hours in which the building is used.
\item Classrooms used in the building and classrooms monitored by GAIA.
\item Number of students and educators occupying the building overall and the classrooms monitored by GAIA. 
\end{itemize}

This step is useful in understanding the potential points of energy consumption in the buildings, and their relative contribution in the energy consumption of the building compared to each other.

\subsection{Step 2 -" Observation and establishing a baseline for energy consumption}

This step involves monitoring the energy consumption of the school building for a time period that is not directly affected by class hours. This will help to identify what is the energy consumption of the building when no class activities take place in it. The following are some examples of options to consider on how to establish this baseline:

\begin{itemize}
\item Days when there is activity inside the school building but no classes take place inside classrooms, e.g., on excursion days.
\item Weekends and national holidays.
\item When the school building is used by other communities after class hours.
\end{itemize}

This is an important step in understanding what part of the energy consumption can be thought of as non-flexible, and which cannot be easily affected by the students and educators when deciding to take specific actions to lower energy consumption inside the building.

\subsection{Step 3 -" Experimentation and monitoring energy consumption in the school during a normal week}

Measure what the energy consumption is during a time period of a ``normal'' week, i.e., a week where no major schedule disruptions take place. E.g., no excursions or other changes to the schools take place, and where the students and teachers do classes as usual. After having established the baseline in the previous step, this will help in identifying:

\begin{itemize}
\item What is the actual percentage of the energy consumption that can affected by the school community, i.e., the part of the total energy consumption that can be targeted without affecting the operation of the school. 
\item Which will be the goals set for lowering the energy consumption and the possible strategies to achieve these goals. 
\end{itemize}

The time period during this step should be at least a week long, and could also take into account the data already available in the system. Steps 2 and 3 help us to identify the portion of energy consumption in which we can intervene. Having identified in step 2 the constant energy needs that are "inflexible" and on which we can not schedule some intervention, we can then calculate the interval between the difference in average consumption and the fixed needs. This is the part of the energy consumption of the building that we can influence, without affecting the orderly operation of the school.

\subsection{Step 4 -" Action to lower energy consumption and monitor the results}

During a time period of at least a week, the school should implement the actions scheduled by the educators to tackle energy consumption, with respect to each cycle of activity chosen by the school. E.g., when the lighting thematic cycle is active, students should implement specific strategies to lower lighting energy consumption. During this period, schools could choose to implement a strategy where they use the tools provided by GAIA to monitor results in energy savings daily, or weekly. The schools could also use strategies in how to implement the activities grouping students in different teams, rewarding them for positive results.

At the end of the period, each school will be able to see the result of such energy-saving actions in its building, and confirm in practice whether these actions will have any impact on the energy consumption of the school.

\subsection{Step 5 -" Staying focused on energy-saving actions and monitor progress}

After having achieved certain energy saving results, schools should focus on continuing to monitor the results and check whether these results persist, or change in some manner. One way to achieve this step is to monitor weekly the respective measurements and reward students and classes based on their progress. Another way is through competitions, e.g., by organizing teams in your school to compete with each other in different parts of the school building. Schools should also have in mind that such aspects are supported within the GAIA competition for this school year.

\section{Application of the Methodology and Results}

In this section, we present an indicative application of the methodology in one of the GAIA schools, in order to explain a bit better how we envision its implementation. We also briefly discuss some other more focused examples of discoveries of existing energy-related issues inside school buildings.

\subsection{A complete run of the methodology - S\"{o}derhamn}

The Staffangymnasiet technical high school in S\"{o}derhamn, Sweden, participates in the GAIA project. It is quite a large school unit, with over 1000 students attending classes, and it has a complete set of sensors installed inside its premises monitoring a good part of its facilities. As a part of its activities in the GAIA project, the school monitors the electrical energy consumption in one its buildings. The thematic cycle chosen by the school for GAIA focused on electricity consumption, attributed mainly to electrical appliances and equipment used during classes. Since this is a technical high school, there are many computers and other related equipment used for a number of hours each day. The school used the methodology as a means to organize an intervention to lower energy consumption engaging mainly the students to act, and the teachers to understand better the patterns of energy consumption and realize the potential impact of long-term interventions. 

\medskip
\noindent\textit{Step 1 - Awareness and Preparatory steps}

The building overall contains eight classrooms, one computer room, one room with 3D printers and laser cutters, three teacher rooms and a couple of small study rooms. Teachers in the school  and mapped class hours that are conducted inside the classrooms in the monitored school building. During an ordinary week, in general the building is used for approximately 140 hours of lecture time. This amount of lecture hours varies depending on things like excursions and visits to external sites.  

\medskip
\noindent\textit{Step 2 - Observation and establishing a baseline for energy consumption} The first thing the school did was to measure how much energy the building consumes when it is not occupied, or in general used for class purposes. They did that during a week of no class, and specifically the 44th week  of the year, which is a holiday period for Sweden. That week the school kept the ventilation running as if it were an ordinary week. The graph below shows the consumption during that week. The higher consumption on Monday is because there was some staff working in the building that day. The reduced consumption on Friday is due to the fact that the school shut down the ventilation one hour earlier on Friday – as is done during a typical school week in most schools. The school building's baseline consumption was calculated from this week. 

\begin{figure}[htbp]
\centerline{\includegraphics[width=0.98\columnwidth]{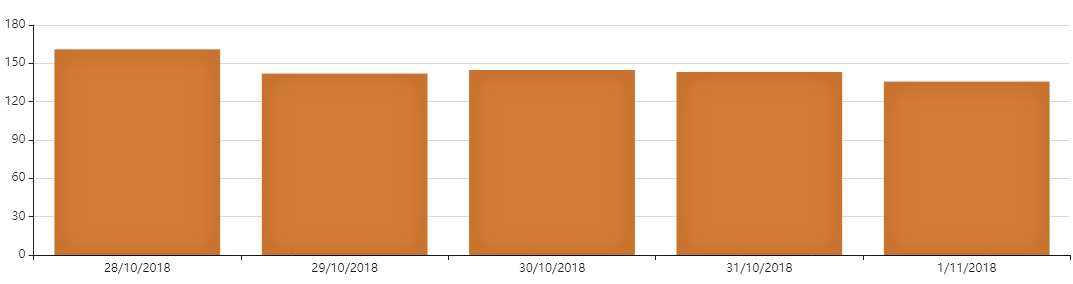}}
\caption{Electric energy consumption week 44 - The baseline week.}
\label{fig:sod1}
\end{figure}

The mean value for this week was calculated as follows: 

\begin{enumerate}
\item Calculating the mean value for Tuesday - Thursday.
\item Approximating the consumption on Monday as the mean value from point 1. This is because the ventilation is turned on the same time during those days. We could not use the real consumption from Monday due to staff working this day.
\item Calculating the mean value for Monday - Friday. 
\end{enumerate}

This was done because the ventilation is turned off earlier on Fridays, and the fact that there was staff working on Monday. The mean value for this week was 141,9kWh/day (calculated as described above). This will essentially be the baseline for this school building.

\medskip
\noindent\textit{Step 3 – Experimentation and monitoring energy consumption in the school during a normal week} The next thing to do was to measure the consumption during a regular week. The best week for that was week 47. During this week, all student groups were on site, with no field trips and absent teachers. Below you can see the consumption during this week:

\begin{figure}[htbp]
\centerline{\includegraphics[width=0.98\columnwidth]{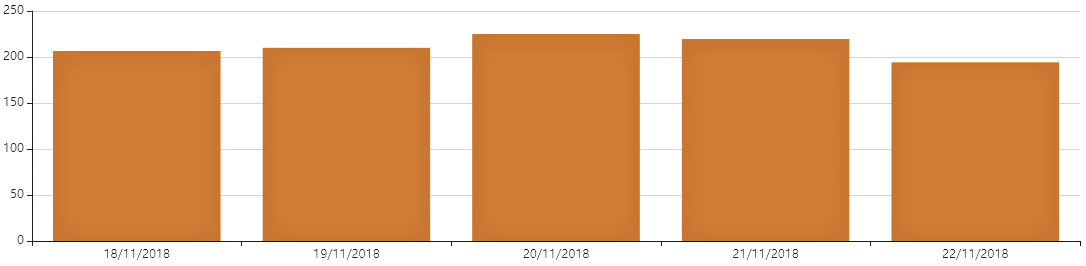}}
\caption{Electric energy consumption during week 47 - The comparison week.}
\label{fig:sod2}
\end{figure}

As you can see, the consumption varies over the week, and that is due to the different amount of lecture time for each day. Below you can see a graph showing the lecture time in the building within the same week:

\begin{figure}[htbp]
\centerline{\includegraphics[width=0.98\columnwidth]{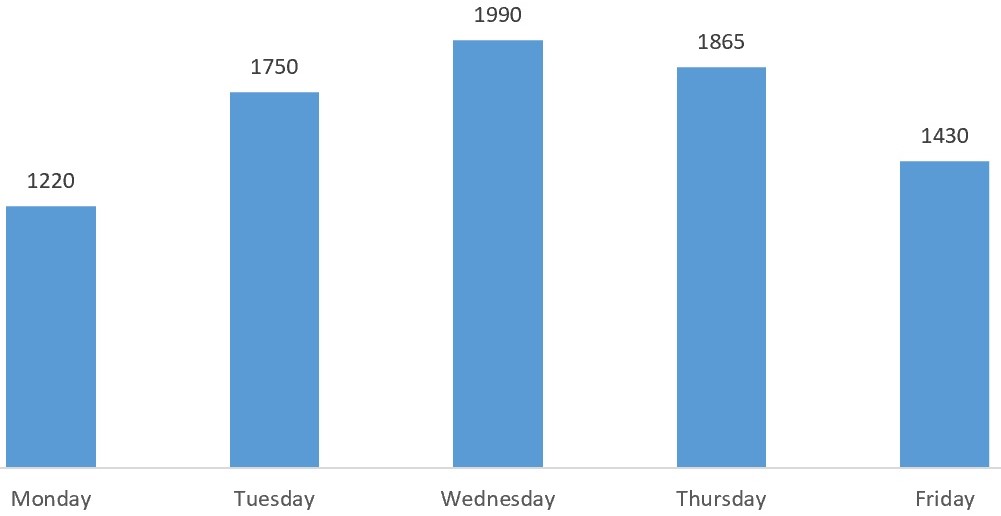}}
\caption{Room usage during week 47}
\label{fig:sod3}
\end{figure}

Consumption clearly correlates with the room usage, with a mean value this week of 211 kWh/day.

\medskip
\noindent\textit{Step 4 – Action to lower energy consumption and monitor the results} Week 50 was the energy saving week for the school in S\"{o}derhamn. This week, all the students and staff were told to turn off electrical equipment when not needed. Several students were monitoring the building and turning off equipment not in use. At the same time, two groups of students competing with each other in the GAIA Challenge game.  Below you can see the consumption for this week:
 
\begin{figure}[htbp]
\centerline{\includegraphics[width=0.98\columnwidth]{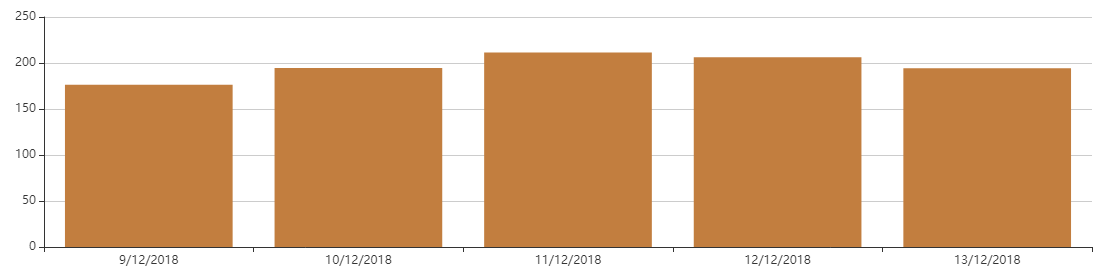}}
\caption{Electric energy consumption during week 50 -" The energy saving week}
\label{fig:sod4}
\end{figure}

The mean value for this week was 196,5 kWh/day. During this week, the school was slightly more occupied than the comparison week (week 47), due to students working until late to finish the preparations for a Christmas show. When we subtract the baseline consumption from the comparison week and the energy saving week, we ended up with a reduction of the energy usage by 21\% during the energy saving week. The difference with week 44 and the baseline, is essentially the part of the energy where the school can intervene. Below you can see a table summarizing some interesting measurements.

\begin{table}
\begin{centering}
\begin{tabular}{lll}
\hline
Week & Consumption & Difference with baseline week \\\hline
Baseline & 141,9 & -\\
Comparison & 211 & 69,1\\
Energy saving & 196,5 & 54,6\\\hline
\\
\end{tabular}
\caption{Power consumption mean values per day (kWh/day)}
\end{centering}
\end{table}

\subsection{Other examples of energy saving opportunities}

\begin{figure}[ht]
\centerline{\includegraphics[width=0.95\columnwidth]{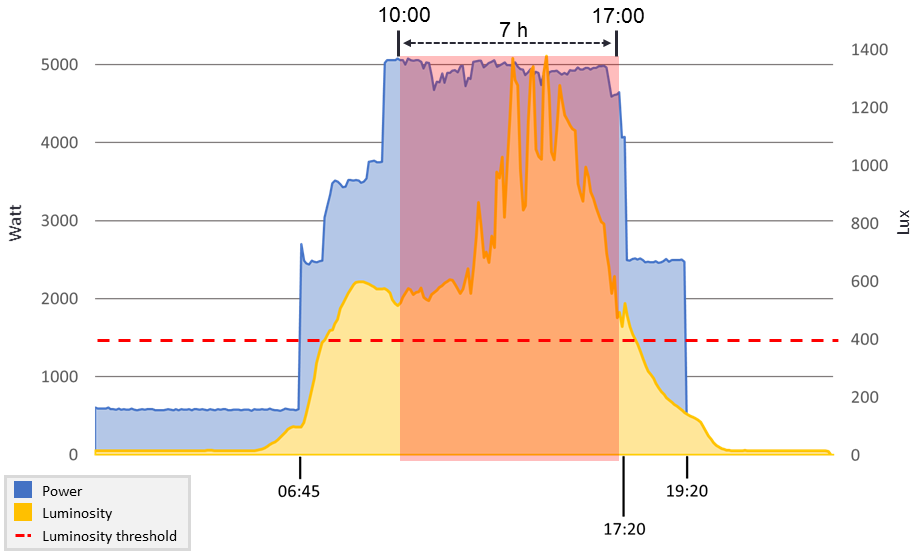}}
\caption{ Graph with light level threshold and with the period in which there is a waste of electricity highlighted}
\label{fig:prato1}
\end{figure}

We proceed now to describe a couple of other characteristic examples of energy-related issues that can investigated using this series of simple steps. Another large high school with close to 1000 students participates in the project. The school decided to focus on the lighting of its halls as the use-case for targeting energy savings. With respect to luminosity, there is a minimum recommended value of 150 lux for an area of passage as an indoor hall. There are a number of luminosity sensors installed in this specific building. Given that the sensors produce that are highly related to their orientation, which is not optimal for calculating a luminosity average value, the students had to approximate the values they saw through the system.  Making a rough estimation, students set a threshold of 400 lux for the values produced by the sensors that they thought it corresponded to ``good enough'' lighting. Figure~\ref{fig:prato1} displays the measurements for power consumed by lights in the hall and luminosity, with the addition of the 400 lux threshold (horizontal line marked in red). Also highlighted in the figure is the interval during which luminosity in the school hall is above the threshold. 

It is evident that between 10:00AM and 17:00PM the lights should be turned off. This is a recurring situation in this specific school building for a number of months, due to its location (Mediterranean) and orientation; i.e., it is not something that is observed only for a single day or over a short time period. The next step was to act on the plant for turning off the unnecessary lights, while also making sure not to leave any part of the hall in the dark. Lighting should be turned off for sufficient time, in order to be able to observe the change in the data. It was convenient to calculate the average values of the lighting system during a ``normal'' baseline period and after the intervention.

The school analyzed the new data regarding power during the period in which the light was turned off. With the lighting configured as usual, power consumption is at approximately 4.9kW. When the school acted to keep active only what is necessary, the power consumption decreased to 1.9kW, thus saving 3kW in the process. This practically means that 21kWh could be saved during a single day, considering the 7 hours of the interval during which this issue was identified.

Another high school in Greece used the GAIA tools and produced some interesting findings in terms of the baseline consumption of its building. Although the school staff were aware of the issue, they were not aware of its scale. More specifically, for a period of 114 days (between January 7 and April 29, 2019) the average total consumption for weekdays was 367kWh, while on weekends it was 119kWh. This essentially means that the school consumed during weekends close to one third of what it did during working days. By focusing on educational activities around energy consumption, they were able to identify the problem, verify its scale and start the process of planning for its resolve.

\section{Conclusions - Future Work}

We presented a methodology for achieving verifiable energy savings following a set of simple steps based on the use of data coming from an IoT infrastructure installed inside school buildings. This methodology, in general, is an attempt at combining a set of specific guidelines with a certain flexibility in terms of choosing the energy domain to focus on, as well as the means to implement a strategy towards energy savings. Overall, the educational domain has many constraints on implementing activities that are not in some way tied to a specific lecture or learning outcome; in this sense, having a well-defined set of tools that provide clear and concise data, and integrating these data within a learning activity can lead to very interesting results.

In terms of actual results from combining this methodology with data produced inside school buildings, we have seen energy saving results in the range of 15-20\% at a number of instances, such as the ones presented in this work. An additional remark is that you don't always need complex tools to achieve good results; in many cases. there is a low-hanging fruit in energy savings in public buildings such as schools, where simple interventions based on actual data can have a real impact. Given the respective consumption data, schools can determine where to focus and adapt their energy-saving strategies. From what we have seen, there is a lot of room for improvement with respect to energy consumption in school buildings, while the human factor plays an important role in achieving tangible energy savings without always requiring other types of intervention, such as building renovation.

In terms of future work, we plan to utilize more sophisticated energy disaggregation algorithms, as well as provide more detailed guidelines and templates for energy-saving activities. We also plan to publish datasets produced from the GAIA infrastructure to help the community work with actual school building data.

\section*{Acknowledgment}

This work has been partially supported by the ``Green Awareness In Action'' (GAIA) project, funded by the European Commission and the EASME under H2020 and contract number 696029, and the EU research project ``European Extreme Performing Big Data Stacks'' (E2Data), funded by the European Commission under H2020 and contract number 780245. This document reflects only the authors' views and the EC and EASME are not responsible for any use that may be made of the information it contains.

\bibliographystyle{IEEEtran}
\bibliography{part-b-bib}

\end{document}